# Composite Fermions and Parton Wavefunctions in twisted Graphene on hexagonal Boron Nitride

J. Salvador-Sánchez[1], A. Pérez-Rodriguez[1], V. Clericò[1], O. Zheliuk[2], U. Zeitler[2], K. Watanabe[3], T. Taniguchi[4], E. Diez[1], M. Amado[1], and V. Bellani[5]

[1]Nanotechnology Group, Department of Fundamental Physics, University of Salamanca, 37008 Salamanca Spain
[2]High Field Magnet Laboratory (HFML-EMFL), Radboud University, 6525 ED Nijmegen, The Netherlands
[3]Research Center for Functional Materials, National Institute for Materials Science, 1-1 Namiki, 305-0044 Tsukuba, Japan
[4]International Center for Materials Nanoarchitectonics, National Institute for Materials Science, 1-1 Namiki, 305-0044 Tsukuba, Japan
[5]Department of Physics, University of Pavia, 27100 Pavia, Italy

**Abstract**

In a twisted graphene on hexagonal Boron Nitride, the presence of a gap and the breaking of the symmetry between carbon sublattices leads to multicomponent fractional quantum Hall effect (FQHE) due to the electrons' correlation. We report on the FQHE at filling factors $v = k/2$ and $v = k/3$ with $v > 1$, and on the composite fermions at in the $v < 1$ lowest landau Level $v = 4/5, 5/7$ and $2/3$. These fractional states can be described with a partons model, in which the electron is broken down into sub-particles each one residing in an integer quantum Hall effect state; partons are fictitious particles that, glued back together, recover the physical electrons. The parton states host exotic anyons that could potentially form building blocks of a fault-tolerant topological quantum computer.

## 1. Introduction

In recent years, significant efforts have been made to understand the topological aspects of the quantum Hall effect [1], particularly in systems such as graphene and graphene based heterostructures [2]. Most studies on Fractional Quantum Hall (FQHE) in graphene on hexagonal Boron Nitride (hBN) superlattices have focused on the lowest Landau level (LL) states. These FQH states observed in two-dimensional systems can be well-described by the composite fermions (CFs) model, where an even number of magnetic flux quanta are attached to each electron, resulting in quantized states at filling factors $v = p/(2mp \pm 1)$, where $m$ and $p$ are positive integers. CFs are topological particles formed by electron-vortex bound states, and the FQHE states can be understood as CFs undergoing the integer quantum Hall (IQH) effect. However, in graphene-hBN heterostructures the graphene symmetry between carbon sublattices can be broken by the hBN substrate, leading to collective electron interactions that affect the CFs and give rise to new states [3-7]. Recently, attention has been focused on more exotic states resulting from weak residual interactions between CFs; these include CF crystals at low LL filling, chiral *p*-wave paired states of CFs for certain even denominator states, and unconventional series of CF crystals [4]. Odd denominator Abelian FQH states in suspended graphene, bilayer graphene, and graphene on hBN are also subjects of intense investigation [3].

In this work we focus on the FQHE at filling factors $v = \eta/3$ with $v > 1$, and on the composite fermions (CFs) at in the lowest landau Level $v = 4/5, 5/7$ and $2/3$, which belong to the series of integral quantum Hall effect (IQHE) of CF electrons at $v = p/(2mp \pm 1)$ and $v = 1- p/(2mp \pm 1)$ for CF holes, where $2m$ is the number of magnetic flux quanta attached to the CF, or vorticity. In our magnetotransport measurements in a fully-encapsulated graphene-hBN heterostructure we observe FQHE at $v = 4/5, 5/7$, and $2/3$, a subset of the integral quantum Hall effect (IQHE) series attributed to CF electrons, expressed as $v = p/(2mp \pm 1)$, as well as $v = 1 - p/(2mp \pm 1)$ for CF holes. Our primary focus alights upon the $v = 2/3$ fraction and its elucidation via a partons' wave function. Within this model, the electron gracefully disassembles into sub-particles, these enigmatic partons, each ensconced within an IQHE state. Intriguingly, these conceptual entities, once intricately reunited, faithfully reconstruct the physical electron. The parton states host exotic anyons that could potentially form building blocks of a fault-tolerant topological quantum computer [6].

## 2. Discussion

The sample consists of a fully encapsulated graphene-hBN heterostructure where the symmetry between carbon sublattices is broken by the effect of the encapsulating hBN flakes provoking the appearance of a sizable energy gap, two ingredients which allow for the formation of multicomponent FQHE [7]. Figure 1a shows an optical microscope image of the final hBN-graphene-hBN device with Cr/Au electrical contacts for the measurement of the longitudinal Rxx and transversal Rxy resistances. The graphene is encapsulated between two hBN layers, originally the same flake broken into two, which were aligned forming a 180º angle between them by using their natural edges (more details in the supporting Information).

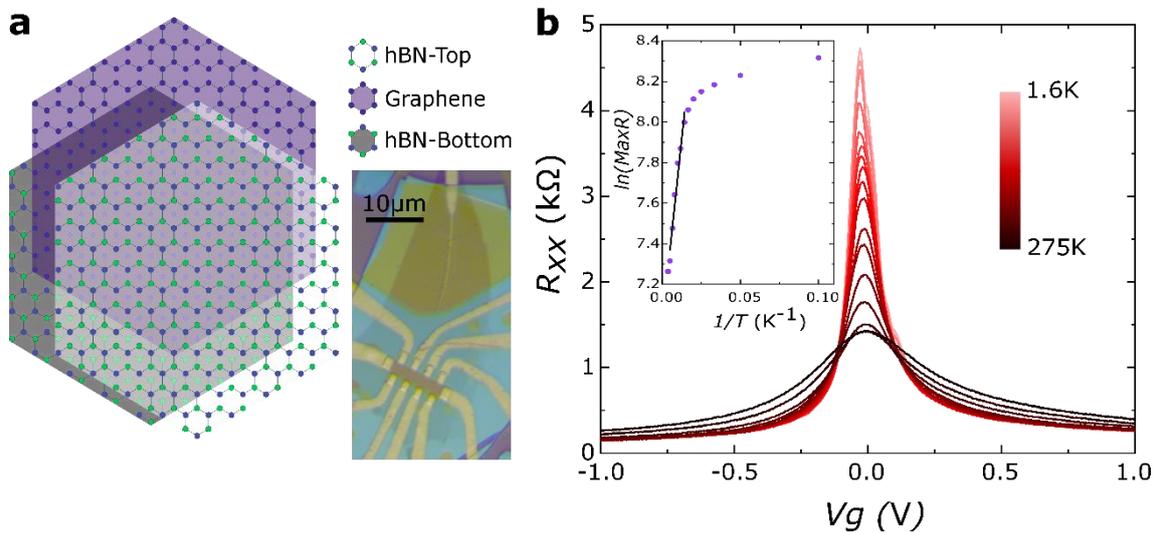

Figure 1: (a) Art view schematic of the lattice configuration with 1.75 ° misalignment between graphene and hBN and 180º misalignment between top and bottom hBN layers and optical microscope image of the fabricated device. (b): Temperature dependence of the Dirac peak in the longitudinal resistance Rxx. The inset shows the temperature dependence of the Rxx Dirac peak maximum value at CNP. From the linear fit we get the energy gap value 12.6 meV.

Transport measurements were taken with the sample placed in 4He cryostat with a base temperature of 1.4 K, at magnetic fields up to 30 T unless stated otherwise, obtained using standard lock-in technique with an ac-driving current of 10 nA at 17 Hz. The longitudinal Rxx measured at zero magnetic field, as a function of electron and hole densities tuned by the gate voltage (Vg) is shown in Figure 1b; we did not exceed negative Vg of -21.6 V and positive of 23.4 V to prevent current leaks and consequent sample breakdown. As previously reported, the stacking of graphene on hBN results in a Moiré superlattice, leading to the emergence of secondary Dirac points. The wavelength of this superlattice is determined by the relative stacking angle between the graphene and hBN lattices [8, 9]. In our sample, the resistance curve clear Moiré satellite secondary Dirac peaks at Vg ~ 20 V, corresponding to charge densities of around $\pm 8.5 \times 10^{12}$ cm$^{-2}$. From the positions of the satellite Dirac peaks we calculate a misalignment between graphene and hBN of 1.75° (see Supporting information). In order to obtain the value of the energy gap, we measured the temperature dependence of the Dirac peak (Figure 1b) and we plotted the value of its maximum as a function of temperature with an Arrhenius plot (inset in Figure 1b). The thermally activated transport exhibits two distinct regimes with an activated behavior at low temperatures and with a maximum value for the Dirac peak of just over 1 KΩ at room temperature which is consistent with the presence of small energy gaps. The value for the gap was obtained from the linear region of $R_{xx}$ as a function of 1/T in an Arrhenius plot resulting in $E_{gap}$ = 12.6 meV. From the low field magnetotransport measurements we can estimate a carrier mobility of 250.000 cm$^2$V$^{-1}$s$^{-1}$ throughout the sample (as detailed in Supporting Information). To further check the quality of the final heterostructure, a thorough Raman characterization was performed prior to the etching process and after the deposition of the metallic contacts by means of a Raman spectroscopy, shown in detail in the Supporting Information. We checked the quality of the graphene layer from the ratio of the intensity of the 2D and G peaks in Raman spectra of graphene, providing a value I(2D)/I(G) = 8. Raman analysis also was used to identify the most homogeneous, impurity-free area where the final device was finally fabricated.

Magnetotransport measurements at moderate temperatures and magnetic fields are displayed in Figure 2 where we report the longitudinal $\sigma_{xx}$ and transverse $\sigma_{xy}$ conductivities.

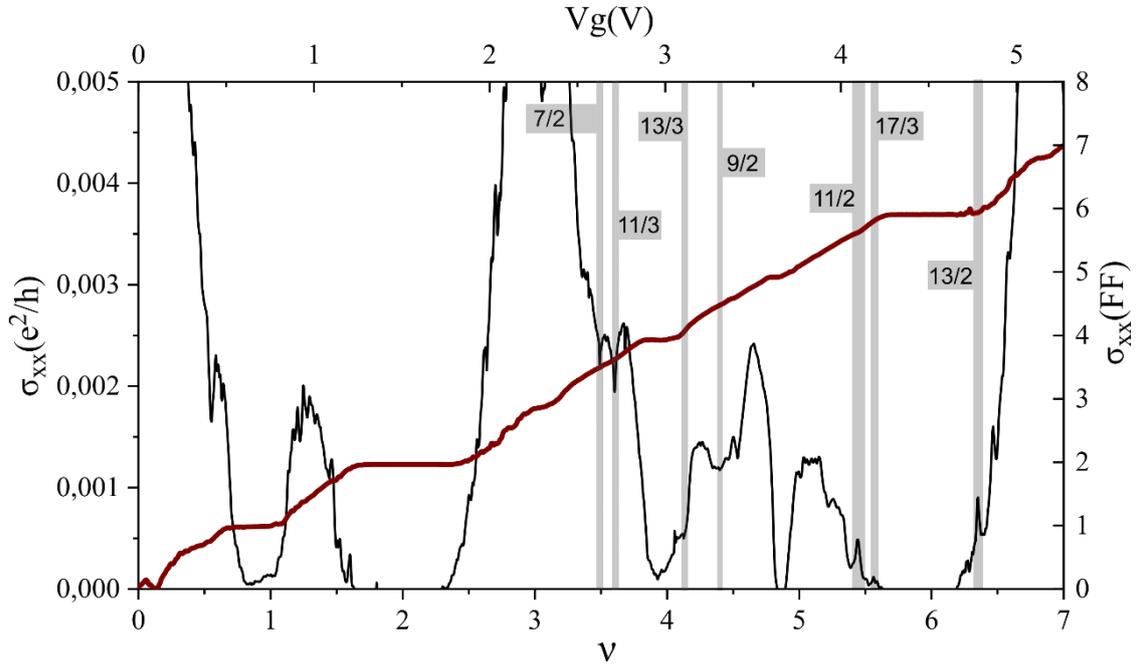

Figure 2: Longitudinal $\sigma_{xx}$ and transverse $\sigma_{xy}$ conductivities as the LLs are gradually filled by tuning the back-gate voltage ($V_g$). Both curves were recorded at 1.4 K with an external fixed magnetic field of 10 T. Several fractions arising from the FQHE can be clearly identified even at such moderate temperatures.

These data were obtained in a He4 cryostat at 1.4 K as a function of $V_g$ and at a fixed external magnetic field o $B$ = 10 T. Besides the standard sequence of plateaus for the QHE at the lowest Landau levels, our results are featuring several fractions at $v = k/2$ and $v = k/3$ being $k$ an integer) beyond those observed in previous experiments in graphene monolayer sandwiched between hBN [8], suspended graphene [10] and in graphene on hBN [11, 12] with similar mobilities and external magnetic fields, revealing the high quality of the device. The magnetotransport measurements as a function of $B$ up to 30 T are displayed in Figure 3a and several fractional states in the lowest LL in the highlighted area are shown in Figure 3b. Our results exhibit a pronounced dip in the longitudinal resistance at magnetic fields corresponding to 2/3 filling which in turn corresponds, alongside with 1/3, to the most robust odd denominators in the FQH states associated with the largest quasiparticle energy gaps [12, 13]. This state is well described by the CF model in which the FQHE for the electrons is interpreted as the IQHE for CFs, quasiparticles originating from the electron mutual interaction in the magnetic field [14, 15]. Moreover, minima associated to the 1/5 and 1/7 fractions are visible around the 2/3 state which is a remarkable observation in such a moderate temperature, highlighting the overall quality of the graphene heterostructure.

In the CF model developed for 2D electrons in semiconductor systems, the CF are a bounded state of an electron with an even number of magnetic flux quanta, and the electron-electron interaction is well effectively incorporated into the CFs themselves. For instance, at magnetic fields around $v = 1/2$, CFs constituted by an electron with two flux quanta are formed ($^2$CF), while at $v = 1/4$ with four flux quanta ($^4$CF). The global result is the transition from a system of interacting electrons in a magnetic field $B$ to another of vanishing interacting CFs in a residual effective field $B^* = B - 2\Phi^0\rho$, where $\Phi^0 = hc/e$ is the flux quantum and $\rho$ the electron of CF density [14]. The change of the magnetic field from $B^*$ leads to the creation of Landau-like levels called $\Lambda$ levels ($\Lambda$Ls) in $B^*$, with a filling factor $v^*$ related to the electron filling factor $v$ by the relation $v = v^*/(2v^* \pm 1)$. The IQHE of composite fermions at $v = n$ manifests as a FQHE at odd-denominator fractions of the form $v = n/(2n \pm 1)$, which can be referred to as the "conventional" FQHE states [16].

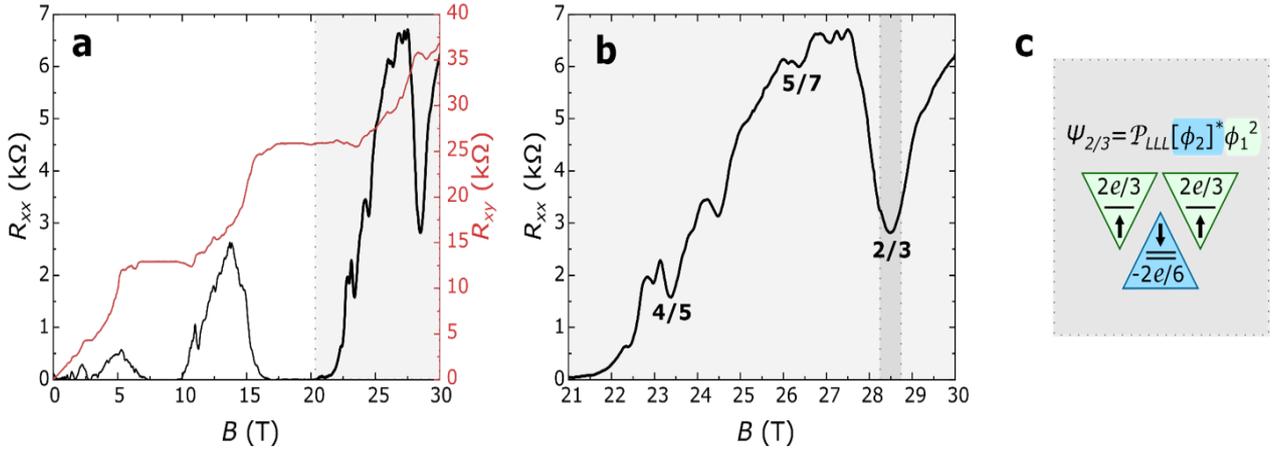

Figure 3: (a) Longitudinal resistance Rxx of our graphene-hBN sample at external magnetic fields up to 30 T. (b) In the highlighted high-field region we can clearly identify three major dips in Rxx, associated with three distinct FQH states. (c) Wave function and schematic view for the parton structure corresponding to the $v = 2/3$ Landau level. Arrows indicate the effective magnetic field seen by each parton, which can be antiparallel to the one experienced by the electrons, while horizontal lines represent the occupied LLs from where each parton arises.

Jain introduced the parton model to generalize the concept of CFs, building a wider class of FQH wave functions that explain these unconventional fractional states [5]. The parton model describes the standard states of CFs, but also more general states that also support non-Abelian excitations. The parton construction gets the standard states of CFs, as well as more general states that do not bring by themselves to an interpretation in terms of composite fermions, such as states that support non-Abelian excitations in graphene and other two-dimensional systems [17]. The usual proposals for non-Abelian anyons and Majorana particles, which are neither fermions or bosons, are related to heterostructures' realization of topological superconductivity in two dimensions. In the parton model one imagines dividing the electron into $q$ fictitious objects called partons and, with them, unusual fractional states can be built, placing each parton in an IQH state with filling factor $n_\delta$ where $\delta = 1, 2, \ldots, k$ labels different parton species [18, 19]. The partons are non-physical objects that should be stuck back together to recover the electrons, and the wave function of the incompressible unconventional FQH state at filling $v$ is described as

$$\Psi_v^{n_1 n_2 n_3 \cdots} = \mathcal{P}_{LLL} \prod_{\delta=1}^{k} \Phi_{n_\delta} \qquad (1)$$

where $\Phi_n$ is the Slater determinant wave function for the state with $n$ filled Landau levels (LLs) of noninteracting electrons, and $\mathcal{P}_{LLL}$ denotes projection into the lowest LL as is appropriate for the high field limit [18, 19]. The effective magnetic field seen by the partons can be antiparallel to that seen by the electrons, and in this case, it corresponds to negative filling; this corresponds to negative fillings for the partons, denoted as $\underline{n}$ with $\Phi_{\underline{n}} = \Phi_n^* = \Phi_{-n}$. Since the density of each parton species equals the electronic density and all the partons are at the same magnetic field, the charge of the $\delta$ parton species is given by $e_\delta = -\nu e/n_\delta$. The constraint that the sum of the parton charges is the electronic charge implies that the electron filling is related to the partonic filling by $\nu = \left[\sum_{\delta=1}^{k} n_\delta^{-1}\right]^{-1}$ [6, 18, 19, 20-23]. Several FQHE states can be reinterpreted with the parton picture. For instance, the Laughlin $\nu =1/(2p+1)$ state is described by the wavefunction $\Psi_{1/(2p+1)}^{Laughlin} = \Phi_1^{(2p+1)}$ and can be seen as the $(2p + 1)$-parton state where each parton forms a $\nu =1$ IQHE state [18, 19]. Within the parton model [24], the wavefunction:

$$\Psi_{2n/(5n-2)}^{\underline{n}21^3} = \mathcal{P}_{LLL}\,[\Phi_n]^*[\Phi_2]^*\Phi_1^3 \qquad (2)$$

for $n = 1$ describes the $\nu = 2/3$ state, and in Figure 3c we show a schematic view of the partons that make up this state. Equ. (2) also describes the $\nu = 1/2$ state ($n = 2$), and the $\nu = 6/13$ ($n = 3$) anti-Pfaffian state which has been observed in bilayer graphene [25]. The Jain states, described by the wave function $\Psi_{\nu = n/(2pn\pm1)}^{Jain} = \mathcal{P}_{LLL}\Phi_{\pm n}\,\Phi_1^{2p}$ are interpreted as $(2p + 1)$-parton states in which $2p$ parton in in the $\nu = 1$ state and one parton is in a $\nu = \pm 1$ IQHE state [18].

Parton states beyond the Abelian Laughlin and Jain ones have been proposed as candidates to describe some FQHE states in higher Landau Levels in monolayer [26] and multilayer graphene [21]. Theoretical and experimental studies focusing on the spin polarization in graphene FQHE states in the lowest and first LLs are underway, exploring their Abelian or non-Abelian nature. Balram et al. [6] found that the FQHE in graphene can be effectively described in terms of composite fermions, although the spin physics in the lowest LL differs qualitatively from that in the first LL. To this regard, recent experiments on the FQHE states in the first LL of graphene have suggested a nontrivial role of the spin degree of freedom [11], stimulating theoretical studies on the competition between fractional quantum Hall states with different spin polarizations [22]. The 5/7 state in the of graphene, which we observe at B ~ 23.4 T, is particularly intriguing; theoretical works have indicated that the fully spin-polarized and spin-singlet states are nearly degenerate in terms of their Coulomb interaction energy [23], while exact diagonalization studies have been unable to determine which of the two states is the ground state. They are instead interpreted in terms of interacting CFs in fractional regime and of the parton model, which was introduced by Jain [5] and more recently used to describe unusual FQHE states [6]. We also focus our study on the 2/5 spin-singlet Jain state, which can be described by the conventional CF model in the IQHE regime; Balram recently described [6] how as the magnetic field ($B$) is lowered, the FQHE state at $\nu = 2/5$ transitions from a spin-singlet CF unpolarized state to a fully polarized one and eventually at low $B$ goes to a parton state. Regarding the $\nu = 4/11$ and 6/17 states, they can also be described as parton states and, to our knowledge, have not been observed in single or bilayer graphene or graphene heterostructures.

The parton model is particularly interesting for these stated due to the FQHE of composite fermions (CFs), which haves have unusual topological order and cannot be explained by the standard series of integral quantum Hall effect (IQHE) of CFs at $\nu = p/(2mp \pm 1)$. Among them we have the states at $\nu = 4/11$, 5/13 and 6/17 which are extraordinarily fragile and unusual, and only been observed in high mobility semiconductor 2D systems [17, 27-31]. The nature of the states at $\nu = 4/11$, 5/13 and 6/17 is still controversial and cannot be understood as either an IQHE or a paired state of CFs [16]. They have already been observed in two-dimensional semiconductors [17, 27, 29], [30, 31] and the 4/11 in graphene [32], and they can be described as FQHE states of CFs, indicating that the CF-CF interaction cannot be assumed to be negligible [16], [27]. In this picture CFs form Landau-like levels called $\Lambda$ levels ($\Lambda$Ls) in $B^*$ and their filling factor $\nu^*$ is related to the electron filling factor $\nu$ by the equation $\nu = \nu^*/(2\nu \pm 1)$. The IQHE of composite fermions at $\nu = n$ manifests as FQHE at odd-denominator fractions of the form $\nu = n/(2n \pm 1)$, which are the conventional FQH states [16]. For the FQHE at $\nu = 4/11$, the CF filling factor is $\nu^* = 1 + 1/3$, and for the $\nu = 5/13$ state, it is $v^* = 1 + 2/3$. The FQHE at $v = 6/17$ corresponds to $v^* = 1 + 1/5$ [16, 27]. Theoretical analyses and numerical simulations by Mukherjee et al. [16] have shown that the mechanism of FQHE at $v = 4/11$ and $5/13$ states is unconventional. These states arise from a filling $v^* = 4/3 = 1 + 1/3$ and $v^* = 5/3 = 1 + 2/3$ of CFs carrying two vortices in the same direction as the external magnetic field (parallel vortex attachment), respectively. However, the $v = 1/3$ and $2/3$ states in the second $\Lambda$L of composite fermions belong to the Wójs-Yi-Quinn (WYQ) type, which is topologically distinct from the conventional Laughlin state [20]. Recent theoretical work has shown that spins play an important role in these states. For example, at $v = 4/11$ FQHE, the CF fermions completely fill the lowest spin-up $\Lambda$L and form a conventional $1/3$ state in the spin-reversed lowest $\Lambda$L, with polarization $\gamma = (v^*_\uparrow - v^*_\downarrow)/(v^*_\uparrow + v^*_\downarrow)$, where $v^*_\sigma$ represents the filling factor of composite fermions with spin $\sigma$. The theory indicates that the $v = 4/11$ state is a partially spin-polarized conventional state that transitions to a fully spin-polarized unconventional state at a critical field with increasing $B$ [16].

These theoretical results, combined with the experimental identification of their incompressibility, i.e. absence of an energy gap in the excitation spectrum [30], open the possibility of a novel topological order. The same process described above for CFs and $\Lambda$ levels can be applied to the $v = 6/17$ state, where the correspondence is $v = 6/17 \rightarrow v^* = 1 + 1/5$, as described by Pan et al. [27]. In the parton model, the FQHE state at $v = 4/11$ is described by the parton state $4\underline{2}1^3$ with wavefunction $\Psi^{4\underline{2}1^3}_{4/11} = \mathcal{P}_{LLL}\Phi_4[\Phi_2]^*\Phi_1^3 \sim \frac{\psi^{Jain}_{4/9}\psi^{Jain}_{2/3}}{\Phi_1}$ [18, 19], where the sign $\sim$ indicates that the states on either side of the sign differ in details of how the projection to the LLL is implemented. As regards the $v = 6/17$ state, its parton state is denoted as $3\underline{2}1^3$, with wavefunction $\Psi^{3\underline{2}1^3}_{6/17} = \mathcal{P}_{LLL} \Phi_3[\Phi_2]^*\Phi_1^3 \sim \frac{\psi^{Jain}_{3/7}\psi^{Jain}_{2/3}}{\Phi_1}$ and this state is the $n = 3$ member of the $n\underline{2}1^3$ sequence, being the $n = 4$ relevant for the $4/11$ state FQHE [19]. Finally, the $2/5$ state is the state denoted as $\underline{2}^31^4$ described by the wavefunction $\Psi^{\underline{2}^31^4}_{2/5} = \mathcal{P}_{LLL} [\Phi_2]^{*3}\Phi_1^4$ [6].

## 3. Conclusions

In conclusion, we have employed high quality hBN-graphene-hBN heterostructures to observe FHQE states at filling factors $v = \eta/3$ with $v > 1$, and on the composite fermions (CFs) at in the $v < 1$ lowest landau Level $v = 4/5$, $5/7$ and $2/3$. We evidenced how the $v = 2/3$ fraction is described with a partons model, in which the electron is broken down into nonphysical sub-particles, the partons, each one residing in an IQHE state, that glued back together recover the electrons. We discussed how the parton model describes the states at $v = 4/11$, $5/13$ and $6/17$ that cannot be explained by the usual model of IQHE of composite fermions but are instead due to the FQHE of interacting CFs. These states observed in the two-dimensional electrons in semiconductors heterostructures 2DEGs, were still elusive in graphene. In these unconventional $4/11$, $5/13$ and $6/17$ FQHE states, the electrons are assumed to be composed of partons in the IQHE regime which are unphysical objects that, glued back together, recover the physical electrons. Recent theoretical work by Wu et al. [33] has shown that bilayer graphene can support a new kind of parton FQHE state at $v = 1/2$, the $\Phi_2\Phi_2\Phi_1$, named also as 221. This state supports fractionally charged excitations with non-Abelian braid statistics but does not represent a chiral pairing of CFs and is topologically distinct from Pfaffian and anti-Pfaffian states. Unconventional pairing in single-layer graphene is also theoretically investigated within a Bardeen-Cooper-Schrieffer (BCS) perspective by Balram et al. [23]. In gapped bilayer graphene recent theoretical works show that the tuning of the external magnetic field leads to topological quantum phase transitions from an Abelian composite fermion state to a non-Abelian parton [6]; these states host exotic anyons that could potentially form building blocks of a fault-tolerant topological quantum computer. Some of these states have been observed in a recent experiment [25], more experiments are needed to deeply understand their quantum nature.

Non-Abelian topological order represents a sought-after state of matter boasting new characteristics, notably the intriguing ability of quasiparticles to retain the order of their exchanges. These unique anyonic excitations hold great promise as fundamental units for constructing fault-tolerant quantum computers. Despite concerted efforts, achieving non-Abelian topological order and its associated excitations has remained challenging, unlike the more readily attainable quasiparticles or defects found in Abelian topological order. A recent work allowed the successful realization of non-Abelian topological order within the wavefunction generated by a trapped quantum processor, with precise control over its anyons [36]. The use of non-Abelian anyons in graphene based heterostructures, could allow to transfer the method from a trapped ion system to a solid-state graphene-based device; quantum Hall effect in graphene holds up to room temperature, and the possibility of manipulating fractional states at temperature above the millikelvin regime is realistic.

**Aknowledgements and data availability**

We thank A. C. Balram for the critical reading of the manuscript and the enlightening discussions. This work has been funded by projects PNRR-PE0000023-*NQSTI*, MICINN-FEDER, *Broxynano* PID2019-106820RB, *Equaset* PID2022-136285NB PGC2018-094180-B-I00, INFN *Quantep* and JCYL *Cunasol* SA106P23. A.P-R acknowledges funding from the European Union's Horizon 2020 research and innovation program under the Marie Skłodowska Curie-COFUND within the USAL4Excellence program (grant agreement 101034371). Part of this work was supported by HFMLRU/NWO-I, a member of the European Magnetic Field Laboratory (EMFL). Data sets generated during the current study are available from the corresponding author on reasonable request.